\documentclass[journal=jpcafh,manuscript=article,layout=twocolumn]{achemso}

\makeatletter
\let\l@addto@macro\relax
\makeatother
\usepackage[fontsize=10pt]{scrextend}

\usepackage{titlesec}
\usepackage{chemformula,hyperref} 
\usepackage[T1]{fontenc} 
\usepackage[version=4]{mhchem}

\author{Michael C. McCarthy}
\email{mccarthy@cfa.harvard.edu}
\affiliation{Center for Astrophysics $\vert$ Harvard \& Smithsonian, 60 Garden Street, Cambridge MA 02138, USA}
\author{Brett A. McGuire}
\affiliation{Department of Chemistry, Massachusetts Institute of Technology, Cambridge, MA 02139, USA}
\alsoaffiliation{National Radio Astronomy Observatory, Charlottesville, VA 22903, USA}
\alsoaffiliation{Center for Astrophysics $\vert$ Harvard \& Smithsonian, 60 Garden Street, Cambridge MA 02138, USA}

\title{Aromatics and Cyclic Molecules in Molecular Clouds: A New Dimension of Interstellar Organic Chemistry}

\begin{document}




\begin{abstract}
Astrochemistry lies at the nexus of astronomy, chemistry, and molecular physics.  On the basis of precise laboratory data, a rich collection of more than 200 familiar and exotic molecules have been identified in the interstellar medium, the vast majority by their unique rotational fingerprint.  Despite this large body of work, there is scant evidence in the radio band for the basic building blocks of chemistry on earth -- five and six-membered rings --  despite long standing and sustained efforts during the past 50 years. In contrast, a peculiar structural motif, highly unsaturated carbon in a chain-like arrangement, is instead quite common in space. The recent astronomical detection of cyanobenzene, the simplest aromatic nitrile, in the dark molecular cloud TMC-1, and soon afterwards in additional pre-stellar, and possibly protostellar sources, establishes that aromatic chemistry is likely widespread in the earliest stages of star formation.  The subsequent discovery of cyanocyclopentadienes and even cyanonapthlenes in TMC-1 provides further evidence that organic molecules of considerable complexity are readily synthesized  in  regions  with  high  visual  extinction but where the low temperature and pressure are remarkably low. This review focuses on laboratory efforts now underway to understand the rich transition region between linear and planar carbon structures using microwave spectroscopy.  We present key features, advantages, and disadvantages of current detection methods, a discussion of the types of molecules found in space and in the laboratory, and approaches under development to identify entirely new species in complex mixtures. Studies focusing on the cyanation of hydrocarbons and the formation of benzene from acyclic precursors are highlighted, as is the role that isotopic studies might play in elucidating the chemical pathways to ring formation.
\end{abstract}



\section{Introduction}

Astrochemistry is a thriving sub-discipline of astronomy.  Driven by the construction of increasing powerful radio facilities during the past 50 years, and the concurrent astronomical discovery of both well-known and highly unusual molecules, primarily via their rotational spectra, we know now with great confidence that the interstellar medium --- the space between stars --- harbors an astonishing rich chemical inventory \cite{mcguire:17}.  The detection and study of molecules in space is not a mere chemical curiosity.  Rather, molecules are significant reservoirs of chemically active elements, and equally importantly are exquisite probes of the chemical and physical conditions of the regions in which they reside.  Although molecular hydrogen (\ce{H2}) is by far the most abundant molecule (by three to four orders of magnitude), it is extremely difficult to observe directly since this homonuclear diatomic only possesses weak quadrupole-allowed transitions in the infrared \cite{sternberg:269}.  As a consequence, and despite considerably lower abundance, polar molecules serve as highly convenient and sensitive tracers of molecular gas.

One of the most remarkable themes to emerge since the advent of molecular astrophysics in the early to mid-1960s is the preponderance of highly unsaturated carbon chains \cite{avery:47}, an unusual structural motif that is generally uncommon on earth.  This trend is partly a result of selection effects: linear molecules have simple, harmonically related rotational transitions and favorable partition functions compared to rings, which makes their rotational spectra easier to identify in the laboratory and in space.  It is also partly due to kinetic factors: ion-neutral and many neutral-neutral reactions are known to proceed rapidly even at very low temperature \cite{smith:231}.  A key pathway is the polymerization of hydrocarbons by reactions with atomic and diatomic carbon \cite{kaiser:858,millar:195,chastaing:170,herbst:168} to form a variety of acetylenic or cumulenic chains, some as long as \ce{HC11N} \cite{loomis:inpress}.  As a point of comparison, excepting very recent discoveries, the 20 astronomical molecules with a comparable number of carbon atoms to benzene $c$-\ce{C6H6}, the simplest six-membered ring, are all acyclic chains.  The vast majority of the more than 110 polar polyatomic ISM molecules fall close to the prolate limit ($\kappa=-1$, i.e. cigar-shaped), a trend that becomes even more pronounced as the the number of heavy atoms in the molecule increases. 

In contrast, terrestrial organic chemistry is dominated by molecules containing five- and six-membered rings, which frequently serve as the building blocks of polymers and many biological compounds.  It is estimated, for example, that nearly 80\% of the roughly 135~M organic compounds registered in Chemical Abstracts Service (CAS) contain one such ring \cite{lipkus:4443}. For this reason, it is deeply paradoxical that no polar derivative of benzene or a conjugated ring had been identified in space prior to 2018, despite long standing and sustained efforts \cite{kutner_search_1980,myers:155,kuan:650,kuan:31,pilleri:1053,lovas:4345} since the discovery of OH radical in 1963 \cite{weinreb:829}.  This paradox is further deepened because much larger organic molecules, polycyclic aromatic hydrocarbons (PAHs) \cite{Chiar:2013bj,Dwek:1997jd} composed of two or more fused benzene rings are widely believed to be responsible for the Unidentified Infrared Bands (UIRs), \cite{allamandola:733} a prominent set of emission features observed at mid-IR wavelengths (roughly from 3 to 13 $\mu$m).  These bands are ubiquitous in our galaxy and external ones, and may account for perhaps as much as 25\% of all interstellar carbon \cite{Tielens:2008fx}.  While until this year no single PAH had been definitively detected at any wavelength in space, there is compelling circumstantial evidence for this class of molecules based on the frequency coincidence of strong infrared emission features with the characteristic C--C, C--H, etc.~vibrations of aromatic rings and $sp^2$-hybridized carbon.  
 
Owing to the lack of observational data in the rich transition region between linear and planar carbon structures, much uncertainty surrounds the connection, if any, between these two apparently disparate reservoirs of carbon.  Although it was widely assumed early on that formation of most molecules would proceed via a bottom-up approach, top-down arguments have gained significant traction in the last decade \cite{berne:401,zhen:l30,berne:a133}, particularly in light of the recent detection of the highly symmetrical fullerenes, \ce{C60}, \ce{C70} \cite{cami:1180}, and \ce{C60}$^+$  \cite{campbell:322}, which appear unlikely to be formed with any appreciable efficiency via a series of step-wise reactions from small molecules. 

\begin{figure}[t!]
    \includegraphics[width=0.5\textwidth]{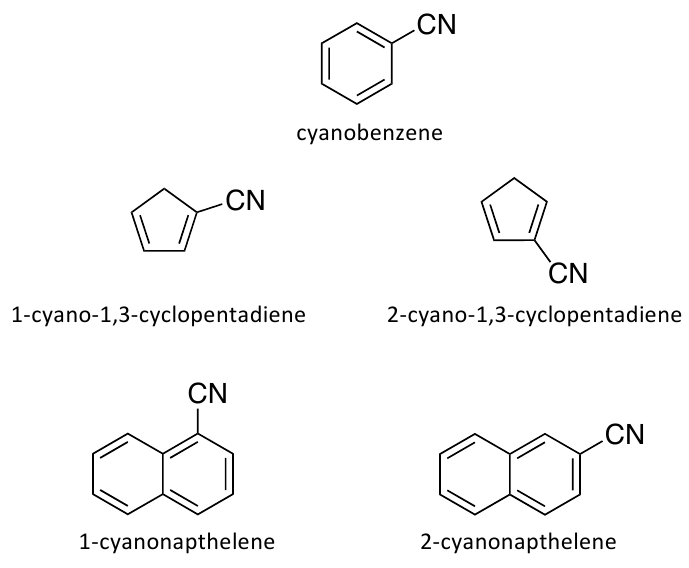}
    \caption{\scriptsize{Structures and common chemical names of the five aromatic or cyclic molecules that have recently been discovered in the dark molecular cloud TMC-1 on the basis of precise laboratory rest frequencies using the 100\,m Green Bank Telescope.}}
    \label{fig:structures}
\end{figure}

Several recent astronomical observations, however, may have begun to shed light on this longstanding conundrum. First,
radio lines of cyanobenzene, $c$-\ce{C6H5CN}, the simplest aromatic nitrile (Fig.~\ref{fig:structures}), were recently observed in the starless core TMC-1 \cite{mcguire:202}.  The Taurus molecular cloud complex is the nearest large star formation region to us, and within it lies TMC-1 which is at a very early evolutionary phase (pre-collapse) characterized by a simple, well-constrained, and homogeneous physical environment. It is a rich source of unsaturated carbon-chain molecules, including cyanopolyynes, acetylenic free radicals, and cumulene carbenes, with many observed in high abundance \cite{Gratier:2016fj}.  Detection in a cold  cloud implies aromatic chemistry is not a phenomenon confined solely to the initial factories of certain carbon-rich stars.  Rather, mechanisms to synthesize at least one small aromatic ring are viable even at extremely low temperature and pressure.  Second, the ease with which this aromatic was subsequently detected in near a half-dozen other molecular clouds \cite{burkhardt:inpress} points to the generality of this chemistry in early star formation, rather than an anomaly specific to one source. 
Third, in rapid succession, evidence was then found for five-membered~\cite{mccarthy:inpress,lee:submitted02} and bicyclic~\cite{mcguire:submitted} CN-functionalized rings in TMC-1 (Fig.~\ref{fig:structures}). Somewhat remarkably, their derived abundances exceed, in some cases by several orders of magnitude, those predicted from chemical models which well reproduce the abundance of various carbon chains irrespective of length, indicating the chemistry responsible for planar carbon structures is favorable but not well understood relative to carbon chains.

\begin{figure}[t]
    \centering
    \includegraphics[width=0.40\textwidth]{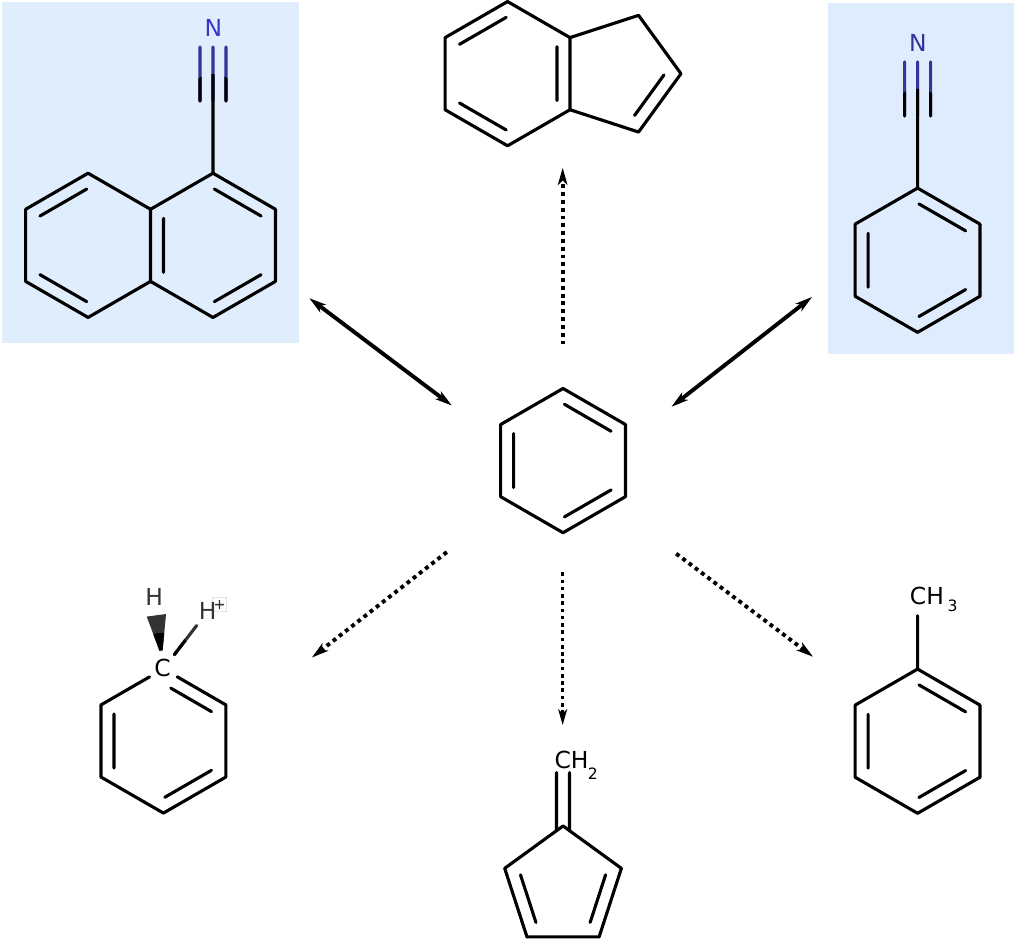}
    \caption{\small Examples of aromatic species that may be of astronomical interest now that cyanobenzene has been found in space.  Radioastronomical evidence for the two molecules highlighted in light blue boxes has now been found.}
    \label{fig:beyond_bn}
\end{figure}

These new findings have opened up an entirely new and largely unexplored area of organic chemistry in space: complex carbon chemistry  in regions with high visual extinction but where the pressure is many orders of lower than terrestrially and the cloud temperature is 10\,K or less. It should be noted that the presence of \ce{CN}-functionalized hydrocarbons such as cyanobenzene in TMC-1 and elsewhere serves as a faithful proxy for the parent hydrocarbon, e.g., benzene, because reactions of unsaturated hydrocarbons with CN, a highly abundant radical in most molecular clouds, are generally exothermic and barrierless~\cite{sayah:177,bennett:8737,Cooke:2020we}.  Furthermore, cyano-derivatives are very favorable targets for radio detection because they have substantial dipole moments and therefore bright rotational spectra.
In turn, this set of discoveries has spawned a multitude of unanswered questions, including what other aromatic molecules exist in molecular clouds (Fig.~\ref{fig:beyond_bn}), what pathways are responsible for their formation, the importance of physical conditions and initial elemental reservoirs in aromatic production, the correlation if any between evolutionary stage and aromatic content, the connection of small aromatics and much larger PAHs, and the chemical and physical role that aromaticity plays in the interstellar medium, among others.  

The purpose of this article is to delve into some of these open questions, with emphasis on laboratory efforts currently underway to better understand the rich and fascinating, but largely unexplored, transition region of carbon from one-dimensional chains to the two-dimensional graphitic structures. The remainder of the paper is organized as follows: Section \ref{methods} summarizes the experimental techniques employed in our laboratory to detect and characterize the rotational spectra of molecules of astronomical interest.  A general discussion of the types of molecules that have been found by these methods is presented in Section \ref{pastwork}, while microwave studies focusing in the formation of benzene and cyanobenzene in our laboratory discharges are highlighted in Section \ref{ring_formation}. Section \ref{new_techniques} is devoted to approaches under development to identify entirely new species in complex mixtures.  The use of isotopic spectroscopy to infer formation pathways under our experimental conditions is given in Section \ref{isotopic}.  Finally, areas of emphasis and future work are presented in Section \ref{futurework}.


\section{Experimental Methods\label{methods}}

\begin{figure}[t!]
    \includegraphics[width=\columnwidth]{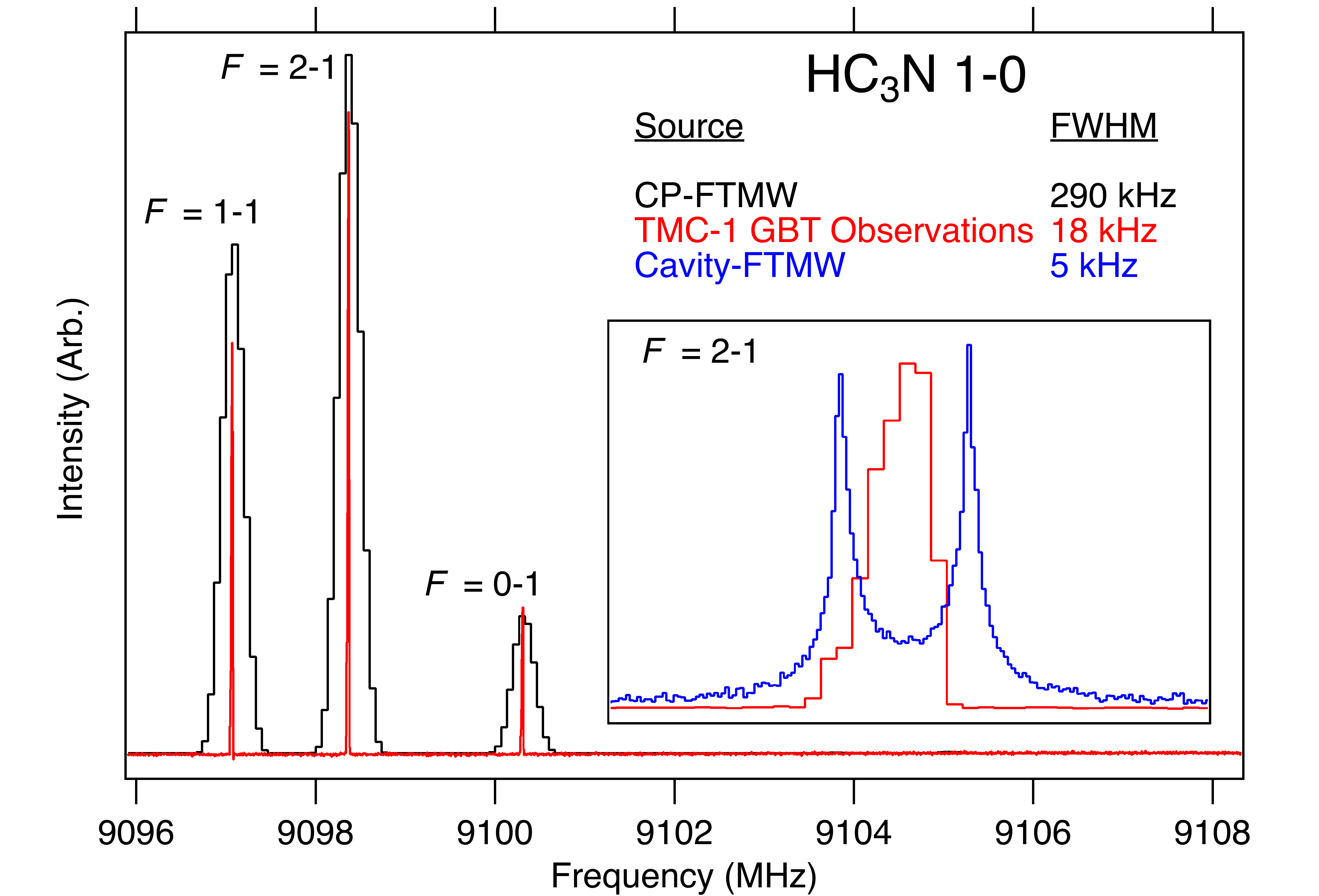}
    \caption{\scriptsize{Laboratory  and astronomical spectra of the fundamental rotational transition ($1 \rightarrow 0$) of \ce{HC3N} near 9.1\,GHz, highlighting the importance of extremely precise laboratory rest frequencies. The astronomical spectrum (in red) was observed toward the cold starless cloud TMC-1 using the 100\,m GBT, and was the result of $\sim$10 hours of on-source integration. Its frequency has been corrected for the characteristic velocity of this source (5.7\,km s$^{-1}$).  The laboratory spectra were recorded with two different microwave spectrometers but used the same molecule source.   The lower resolution spectrum (in black) was recorded with a chirped spectrometer while the higher resolution one (in blue) is with a cavity-enhanced spectrometer.  Because the gas expands along the axis of the Fabry-Perot, each hyperfine line has a double-peaked line shape in the cavity experiment, the result of the Doppler shift of the Mach 2 molecular beam relative to the two traveling waves that compose the confocal mode of the Fabry-Perot. The rest frequency of a transition is simple the arithmetic average of the two frequencies. In the cavity measurement \ce{HC3N} was generated using a pulsed value equipped with a discharge stack, in which a voltage of 900\,V was applied to a dilute gas mixture of benzene and vinyl cyanide in Ne; the spectrum is a concatenation of several 0.2\,MHz scans each of which is the result of 40\,s of integration. The chirped spectrum was recorded using \ce{HC3N} and butadiene as precursors, although only a small portion of the 10\,GHz spectrum is displayed here.} }
    \label{linewidth_comparison}
\end{figure}

The discovery of new molecules in space is highly dependent on the availability of precise laboratory rest frequencies, i.e.~where both the source and the observer share the same rest frame, and hence there is no relative motion between the two. 
On the basis of the close frequency agreement of previously unassigned radio emission lines with those measured (but occasionally calculated), the existence of a new species can generally be made with great confidence.  This situation arises first because rest frequencies can be determined to an accuracy of one part in 10$^6$ or better (e.g.,~to a frequency uncertainty of $\leq$10\,kHz at 10\,GHz) in the laboratory.  The robust matching of these precise frequencies to interstellar features is then aided by the sharpness of radio emission features; in quiescent regions free from star formation, such as TMC-1, linewidths below 0.3\,km sec$^{-1}$ (equivalent to 20\,kHz FWHM at 20\,GHz) are common,\cite{mcguire:202} as Fig.~\ref{linewidth_comparison} illustrates.  In particularly rich astronomical objects, such as the active high-mass star-forming regions Orion and Sgr B2(N), however, lines are both broader (due to turbulence) and more numerous resulting in significant spectral line confusion, a factor that can make new discoveries challenging.  Although many secure molecule identifications have been made in these types of sources \cite{Belloche:2008jy,Belloche:2009ki,Muller:2016kd}, the confidence of others has also been questioned \cite{snyder:914,muller:A92}.  Observations made in colder, more quiescent regions like TMC-1 alleviate many of these concerns.  Not only are the linewidths much narrower but the line density is far lower, i.e.~at centimeter wavelengths, the density is one line (5$\sigma$) per 48\,MHz where a typical linewidth is only 20\,kHz \cite{mcguire:l10}, which corresponds to a filling factor of less than 0.1\% on average.

Laboratory identifications of new species of astronomical interest are frequently made using Fourier transform (FT) microwave spectroscopy paired with a supersonic jet source at centimeter wavelengths \cite{Balle:1981ex,mccarthy:105,mccarthy:611}.  This combination is generally more sensitive than traditional direct-absorption spectroscopy, and when used with an electrical discharge or laser ablation source, has proven highly versatile in generating many astronomical species, including ones much larger than the precursors \cite{thaddeus:757}.  Furthermore, its frequency range ($\leq$50\,GHz) is very well coupled to the low rotational temperature routinely achieved in a jet source (T$_{\rm rot}\leq$5\,K), and at low frequency, searches covering the expected uncertainty (a few percent) of a quantum chemical calculation are highly efficient, i.e.~they can be completed in only a few hours. High-energy isomers are also commonly produced in considerable abundance in discharge sources since collisional cooling efficiently stabilizes these conformers before they can undergo unimolecular isomerization or dissociation.  Because the relative abundances of molecular conformers and isomers in the ISM are rarely thermodynamically controlled,\cite{neill:153,Loomis:2015jh} having access to these potential interstellar species is highly beneficial.

The cavity variant of FT microwave spectroscopy is a well-established measurement technique that dates back to the early-1980s \cite{Balle:1981ex}. It utilizes a large, high-finesse etalon to increase the effective pathlength, albeit at the expense of instantaneous frequency bandwidth (IFBW).  Frequency agility is achieved by computer control, in which individual scans, each about 0.1-0.5\,MHz, are stitched together to produce a seamless spectrum. For this reason, this variant has been used almost exclusively for targeted searches, in which evidence is sought for rotational lines of a specific postulated species on the basis of a theoretical calculation of its molecular structure.  In some instances it has been also used to reproduce a rotational line that exactly agrees in frequency with an unassigned astronomical feature or sequence of lines, i.e.~$B1377$ \cite{Kaifu:2004tk}, the carrier of which was shown to \ce{C6H^-}.\cite{mccarthy:l141}  Owing to high detection sensitivity, this variant is also extremely well-suited for isotopic studies \cite{mccarthy:124304}, and detection of highly-reactive intermediates, such as molecular ions, that are produced in low steady state abundance.\cite{mccarthy:10} Although wide spectral surveys are possible, these become inefficient with increasing IFBW.  As a consequence of limited frequency coverage, it is often very challenging to identify unassigned spectral lines using this narrowband variant, even if the lines are very intense, without a specific prediction of the carrier.

The advent of chirped variant of FT microwave spectroscopy, developed a decade ago by Pate and co-workers \cite{Brown:2008gk} and driven by rapid advances in high-speed electronics, is a complimentary measurement method, in which very wide IFBW -- of order 10\,GHz or more -- is achieved either in a single acquisition or in a series of smaller frequency segments.\cite{Neill:2013bw}  In either version, a pulse-amplified linear frequency chirp is broadcast between a pair of matched microwave horns, and the subsequent free induction decay is digitized in real time by a high-speed, high-sampling rate digital oscilloscope.  Although this variant lacks the sensitivity of the cavity-enhanced version, its wide frequency coverage allows simultaneous detection of rotational lines of multiple species, and its relatively flat instrument response function enables chemical inventories to be derived accurately~\cite{Brown:2008gk}.  One of its most useful astronomical applications is reaction screening, in which wide band spectra acquired using different precursor combinations and discharge conditions are directly compared to astronomical surveys, such as the Prebiotic Interstellar Molecular Survey (PRIMOS) towards Sgr B2 \cite{neill:153} to identify frequency coincidences, and ultimately to identify common carriers.  This approach was used several years ago to identify E-cyanomethanimine ($E$-HNCHCN) \cite{zaleski:l10} and the $E$- and $Z$-isomers of ethanimine in the PRIMOS observations toward Sgr B2(N) \cite{loomis:l9}.  

Although FT microwave techniques have been used with success in providing precise rest frequencies for many postulated astronomical molecules, it should be emphasized that these methods have several limitations.  First, despite the high sensitivity of the cavity variant, more sensitive detection methods exist in the radio band.  Most of these exploit action spectroscopy using a cryogenically-cooled ion trap and mass spectrometric detection in which ion counts of several thousand (or more) are routinely generated.\cite{Brunken:2014cf} Second, rotational spectra of light species, or somewhat heavier ones that only possess $b$-type transitions, may not lie at centimeter wavelengths where  standard cavity-enhanced FT spectrometers commonly operate, although several instruments that perform at the lower end of the millimeter-wave band (e.g., 100\,GHz) have been constructed \cite{kim:296,halfen:ewh07}. Third, the laboratory measurements are almost always fragmentary or incomplete.  Despite high spectral resolution (0.1-1 parts in 10$^6$), owing to limited frequency coverage it is not uncommon that only a combination of rotational constants, a few of the leading centrifugal distortion terms, or both can be derived from the available data.  For this reason, subsequent measurements at millimeter or sub-millimeter wavelengths are often necessary.


\section{Past Studies\label{pastwork}}

\subsection{Linear chains and highly prolate asymmetric tops}

\begin{figure}[t!]
    \includegraphics[width=0.48\textwidth]{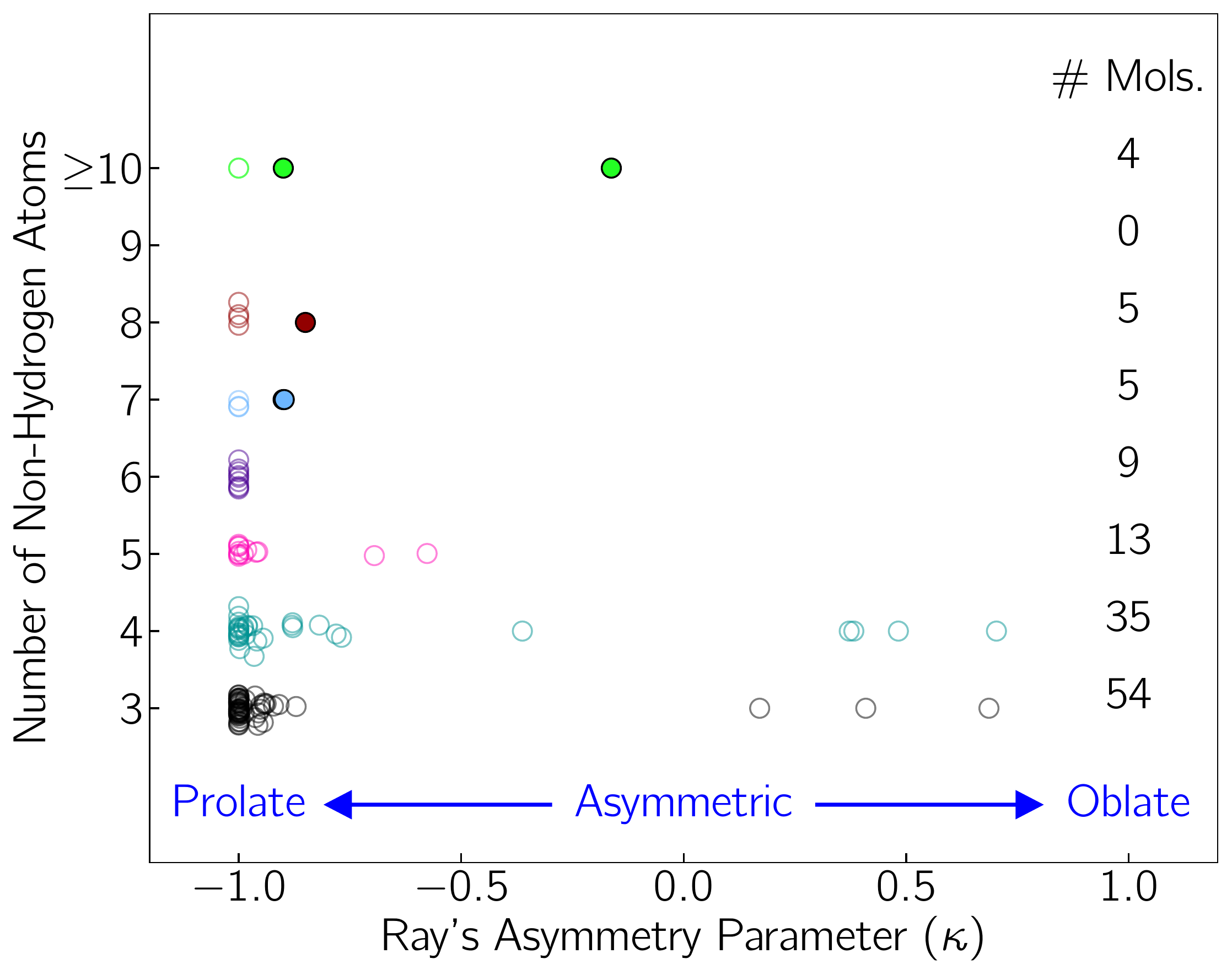}
    \caption{\scriptsize{ Plot of the Ray's asymmetry parameter ($\kappa$, a parameter that describes the degree of asymmetry of a rotor)\cite{Ray:1932yd} for polyatomic, polar interstellar molecules with three or more heavy (non-hydrogen) atoms.  A strictly prolate molecule falls at $\kappa$~=~-1, whereas an oblate one lies at $\kappa$~=~1.  The number of molecules in each group is indicated to the right of each distribution. Compiled data freely available as Supplemental Information to McGuire 2018\cite{mcguire:17} at \url{https://github.com/bmcguir2/astromolecule_census}.} and updated to include molecules detected through 1 Jan 2021.}
    \label{fig:kappa}
\end{figure}

Until quite recently, more than 90\% of known interstellar molecules were linear, prolate, or very near-prolate tops.\cite{mcguire:17} Discounting diatomic molecules, which are prolate by definition, only a handful of even the smallest molecules approach the oblate limit, and all molecules with more than five heavy (non-hydrogen) atoms are prolate or nearly so (see Fig.~\ref{fig:kappa}).  As denoted by the solid circles in Fig.~\ref{fig:kappa}, the five new rings  deviate substantially from the strictly prolate limit, possessing $\kappa$ values in the $-$0.900 (1-cyano-1,3-cyclopentadiene) to $-$0.162 (1-cyanonaphthalene) range.

As in space, laboratory molecular astrophysics studies have frequently focused on strictly linear chains or molecules with nearly linear heavy atom backbones.  This emphasis reflects not only the astronomical importance of this reoccurring structural motif, but several other factors: the accuracy with which rotational constants of long chains can be calculated theoretically (approaching 0.1\%  in the most favorable cases, i.e. chains with closed-shell ground states) or extrapolated from shorter ones; the large dipole moments these chains often possess, a key factor aiding detection in the radio band; and the ease with which harmonically- or nearly-harmonically-related lines can be identified among of forest of unassigned features.  Examples from our laboratory include cyanopolyynes as long as \ce{HC17N},\cite{mccarthy:l231,mccarthy:611} acetylenic radicals up to \ce{C14H},\cite{gottlieb:5433} and cumulene carbides up to \ce{H2C10},\cite{apponi:357} along with various carbon chains terminated with a heteroatom, such as  \ce{SiC8},\cite{mccarthy:766}  \ce{C9S},\cite{gordon:311} and \ce{HC8S}.\cite{gordon:297}  As an example, rotational spectra of \ce{HC19N} and \ce{HC21N} have not yet been measured, but highly accurate estimates of their rotational constants are available (2--4\,kHz) from the extrapolation of the shorter members in the series.\cite{mccarthy:611}  

Laboratory astrophysics historically had lagged radio astronomy in the detection of reactive species, but with the wave of laboratory discoveries starting in the mid-1990s \cite{iida:l45,chen:7828,mccarthy:105,mccarthy:611},
laboratory work is now well ahead, as it should be so that astronomical searches can be undertaken with confidence and efficiently on the basis of precise rest frequencies.  Given the power and versatility of present techniques,
we assert that the rotational study of carbon chains is now largely a solved problem, in the sense that detection of still other chains should be straightforward, provided there is reason to believe that they might be good candidates for astronomical detection.  By inference, it then follows that any new carbon chain detected in space can very likely be detected in the laboratory with present instrumentation or a reasonable refinement of it.
In our laboratory alone, well in excess of 100 carbon chains have been characterized at high spectral resolution  in the past 25 years, with roughly 10\% of these subsequently identified in space to date.




Although dominated by chains and highly-prolate asymmetric tops, a modest number of small cyclic molecules of astronomical interest have also been characterized in the laboratory, but here too selection effects favor detection of those that fall close to the prolate limit.  Examples from our laboratory include two rhomboidal isomers of \ce{SiC3},\cite{apponi:3911,mccarthy:7175} $c$-\ce{C5H},\cite{apponi:l65} and roughly half-dozen ring-chain isomers \cite{travers:l135,mccarthy:l139,mccarthy:305}.

Even though entropic arguments favor the formation of chains relative to rings at high temperatures, owing to the larger density of states that result from the low-lying bending modes of the linear isomer \cite{watts:8372}, studies of multiple conformers of \ce{SiC3}\cite{mccarthy:7175,mccarthy:766} and \ce{C5H2} \cite{mccarthy:518,gottlieb:l141} suggest there is a fairy close correlation between stability and abundance over a wide range of experimental conditions. For this reason, there appears to be no fundamental obstacle  --- kinetic or thermodynamic --- to formation of any isomer, regardless of how unusual its structure might be; rather detection would simply appear to be an issue of sensitivity.
Perhaps the most vivid example in this regard are the four singlet isomers of HCNO.  On the basis of a high-level quantum chemical calculation \cite{schuurman:11586}, isocyanic acid HCNO is predicted to be the most stable isomer arrangement followed by cyanic acid HOCN (24.7\,kcal~mol$^{-1}$), fulminic acid HCNO (70.7\,kcal~mol$^{-1}$), and then isofulminic acid HONC (84.1\,kcal~mol$^{-1}$).  Despite this large energy difference, abundances relative to HNCO of 12\% for HOCN, 2\% for HCNO, and 1\% for HONC have been reported in a jet expansion on the basis of rotational line intensities \cite{mladenovic:174308}.


\subsection{Six-membered rings and small PAHs}

Two reactive six-membered rings have been characterized by FT microwave spectroscopy in our laboratory. Strong lines of both phenyl radical \ce{C6H5},\cite{mcmahon:l61} the prototypical $\sigma$-bonded aromatic radical, and \textit{ortho}-benzyne \ce{C6H4}\cite{kukolich:4353} have been observed using either benzene or a closely-related derivative (e.g., \ce{C6H5Br} or \ce{C6H4Br2}) as a precursor, almost certainly because C--H bond cleavage occurs facilely without rupturing the cyclic carbon skeleton.  Experimental structures for both rings followed \cite{kukolich:2645,martinez:1808}, on the basis of measurements of their singly-substituted isotopic species using precursors enriched in either D or $^{13}$C.  

Rotational spectra of three small, weakly polar PAHs,  acenaphthene (\ce{C12H10}), acenaphthylene (\ce{C12H8}), and fluorene (\ce{C13H10}) have also been observed with the same spectrometer.\cite{thorwirth:1309} Since each sample is a  solid at room temperature, it was necessary to use a nozzle heated to 100-130$^{\circ}$\,C prior to supersonic expansion.  Other PAHs or their derivatives studied by similar methods include corannulene \cite{lovas:4345} and  nitrogen-substituted heterocycles \cite{McNaughton:2008dg}, among others. 

\section{Formation of organic rings\label{ring_formation}}

Despite the large amount of work on carbon molecules in the radio band, very few high-resolution studies have delved into the formation of 5 and 6-membered rings starting from acyclic hydrocarbon precursors, particularly acetylene and diacetylene, which are known to produce long hydrocarbon chains. The lack of studies is in part due to the difficulty of detecting the most stable ring product, non-polar benzene, but also it is one of emphasis: the apparent absence of polar aromatics and cyclic molecules in space.

The astronomical detection of cyanobenzene, however, fundamentally alters this view, demonstrating that synthetic pathways to organic rings are apparently operative even at extremely low temperatures (10\,K) and densities ($\leq10^4$ cm$^{-3}$) that characterize a cold, starless dark cloud (TMC-1). In this source, cyanobenzene could form in two steps:  via the reaction between \textit{trans}-1,3-butadiene and the \ce{C2H} radical to produce benzene (plus H atom), followed by reaction with CN radical. Both reactions have been the subject of extensive theoretical and experimental work \cite{Balucani:1999it,Woon:2006ce,Trevitt:2010ep,Jones:2011yc,landera_mechanisms_2010,Lockyear:2015dn,Cooke:2020we}, and have been shown to be barrierless and highly exothermic.  However, it should be emphasized that the importance of the gas-phase route to benzene in TMC-1 is still a subject of some debate. First, this reaction involves 1,3-butadiene, which is not yet known in space.  The lack of astronomical evidence for 1,3-butadiene itself is not particularly surprising since it is non-polar and its vibrational spectrum possess few distinguishing characteristics relative to other hydrocarbons.  Although still preliminary, however, the apparent absence of its highly polar cyano derivative 1-cyano-1,3-cyanobutadiene ($anti$-($2Z$), \ce{C5H5N}) towards TMC-1 is more concerning. Provided fulvene \cite{baron:1972pa}, a polar, five-membered ring, and the second most stable singlet \ce{C6H6} isomer after benzene, could be detected, it would buttress support for the gas-phase route since this isomer is thought be an important by-product of the reaction \cite{Lockyear:2015dn}.  Second, the abundance of cyanobenzene is well in excess of the latest model predictions\cite{burkhardt:inpress}, implying that benzene is present in greater quantities than can explained. This disparity may indicate other pathways to benzene, such as hydrogenation on grains followed by liberation into the gas phase, may need to be invoked.

Motivated by this astronomical discovery and its chemical implications,
it is obligatory to systematically re-measure the centimeter-wave spectra of many small polar aromatic rings, and to investigate the production of organic rings from acyclic hydrocarbons at high spectral resolution. The transition frequencies of a number of stable aromatic rings were measured many years ago in large gas cells, rather than in supersonic molecular beams.  As a consequence, the line widths of the earlier measurements are much wider than those in TMC-1, resulting in uncertainties in the rest frequencies that correspond to about 10 line widths in this source (see Fig.~\ref{linewidth_comparison}).  

\begin{figure}[t]
    \centering
   \includegraphics[width=0.5\textwidth]{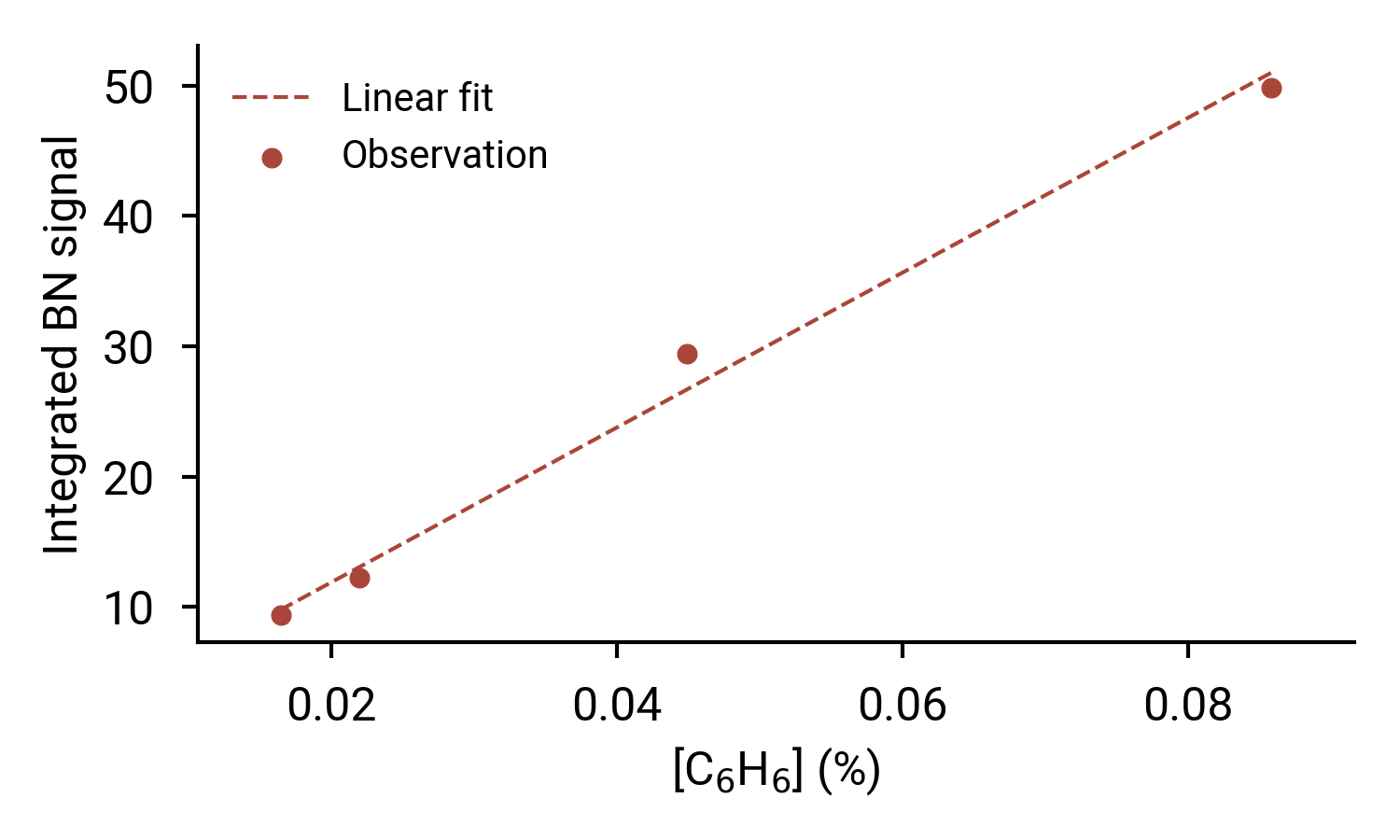}
    \caption{\scriptsize{Experimental relationship between cyanobenzene production and the concentration of benzene in a gas mixture of benzene + \ce{CH3CN} heavily diluted in neon that was subjected to an electrical discharge and a supersonic jet expansion.  The cyanobenzene signal is the integrated intensity of the hyperfine-split $6_{0,6}\rightarrow 5_{0,5}$  transition at 15952\,MHz; estimated uncertainty for each intensity measurement is approximately 20\%.}}
    \label{fig_kinetics}
\end{figure}

A more challenging endeavor is rotational studies of the acyclic-to-cyclic transition, but subjecting  acyclic gas precursors, including one containing the nitrile group (e.g., \ce{CH3CN}), to an electrical discharge, has now yielded evidence for cyanobenzene \cite{lee:2946}  Albeit weak, lines of cyanobenzene were even observed using  hydrocarbons as simple as acetylene.  Although conjectural at this point, it appears likely that cyanobenzene production in this study proceeds via benzene since evidence is found for the fulvene isomer  under the same conditions, and because the production of cyanobenzene obeys first-order kinetics (Fig.~\ref{fig_kinetics}).  These new findings are consistent with earlier work of both \citet{Jones:2011yc} and \citet{Lockyear:2015dn}, and recently by \citet{Cooke:2020we}, although it should be emphasized that the experimental conditions in these studies are markedly different with respect to temperature, pressure, and density. Taken together, these findings establish that cyanobenzene can be used as a tracer for benzene in the presence of CN radical, either in space or in the laboratory \cite{lee:2946}.

In a closely-related study, intense lines of pyrrole \ce{C4H4NH} are produced in an electrical discharge of butadiene and ammonia \cite{lee_interstellar_2019}, suggesting that cyclization may be a general phenomenon under our experimental conditions. Although fragmentary, butadiene appears to be a key species in the ring formation because reaction with a reactive fragment can yield five and six-membered cyclic molecules in a single step, as might be expected by analogy to the well-known Diels-Alder reaction.  The exact nature of this transition however, remains unclear, but should be amenable to the right combination of experiment, theory, and simulation.

The reason benzene is of central importance is formation of the first aromatic ring is the rate-limiting step in PAH growth \cite{Cherchneff:1992}.  Once formed, production of still larger rings is straightforward, although here too the precise mechanism remains the subject of debate, with no fewer than four proposed to date:  phenyl addition/cyclization (PAC) \cite{constantinidis:29064}, methyl addition/cyclization (MAC) \cite{shukla:534}, the popular hydrogen abstraction acetylene addition (HACA) \cite{frenklach:887}, and most recently, clustering of hydrocarbons by radical-chain reactions (CHRCR) \cite{johansson:997}.

\section{Exhaustive complex mixture analysis\label{new_techniques}}

A major experimental obstacle to exploring the rich transition region from chains to rings is the difficulty of identifying entirely unknown species in complex mixtures of familiar and exotic molecules, many of which may not have been the subject of prior calculation or experiment, and may not possess a distinctive spectroscopic signature.  Nevertheless, if this challenge can be overcome, many important questions might then be answered, e.g.,~at what size are rings more stable than chains, what are the dominant ring structures, what are the key intermediates and mechanisms that are responsible for ring formation, among others.

To help answer these questions, it is necessary to rapidly and exhaustively analyze complex chemical mixtures which might consist of many tens or even hundredes of compounds, often of uncertain stoichiometry and unknown structure.  This scenario is of course routinely encountered when seeking to identify a specific new molecule of astronomical interest in the laboratory, because few, if any, synthetic methods are available to selectively produce most species, owing to their transient nature.  As a consequence,  non-selective production methods, commonly electrical discharges, have been used  with considerable success by our group and others. In this scenario, evidence is sought for a new molecule with the reasonable expectation that many other species are simultaneously produced, often in higher abundance.  However, any such lines are discarded or ignored in the analysis, regardless of intensity, if they are not rigorously consistent with the predicted pattern sought.  The difference and challenge from this scenario is we now seek to perform spectral analysis in a more general and unbiased way, with little or no emphasis on a specific target species. 

The technique of microwave spectral taxonomy (MST) by Crabtree \textit{et al.}~\cite{crabtree:124201} seeks to address this challenge.  Its strength is that it combines the chirped and cavity variants of FT microwave spectroscopy to quickly acquire spectra over a larger portion of the centimeter-wave band, and an efficient but exhaustive spectral analysis is then undertaken by only focusing effort on the relatively small number of observed features in a wide spectral range (i.e.~relative to the much larger number of available resolution elements).
 A key component of this approach is the use of automated double resonance to identify lines that arise from a common carrier, and therefore must share a common upper or lower rotational level \cite{martin:124202}.  When combined with an efficient, high-speed program for the prediction and fitting of rotational spectra \cite{carroll:jms} by comparing  frequencies of these small number of linked lines with several billion possible synthetic spectra, it is often possible to determine a unique set of rotational constants.  Because these tools are empirical and unbiased, taken together they can be used to identify highly abundant molecules that have never been the subject of prior experimental or theoretical study.  Examples include \ce{HNSiN}, \ce{H3NSiN} \cite{crabtree:11282}, and \ce{HC4C(O)H} \cite{mccarthy:154301}, among others.  A recent MST analysis of the discharge chemistry starting from diacetylene and carbon disulfide,  for example, resulted in the detection and assignment of more than 85 unique variants, including several new isotopic species and more than 25 new vibrationally excited states of \ce{C2S}, \ce{C3S}, and \ce{C4S}.\cite{mcguire:13870}
 
\begin{figure*}
    \centering
    \includegraphics[width=\textwidth]{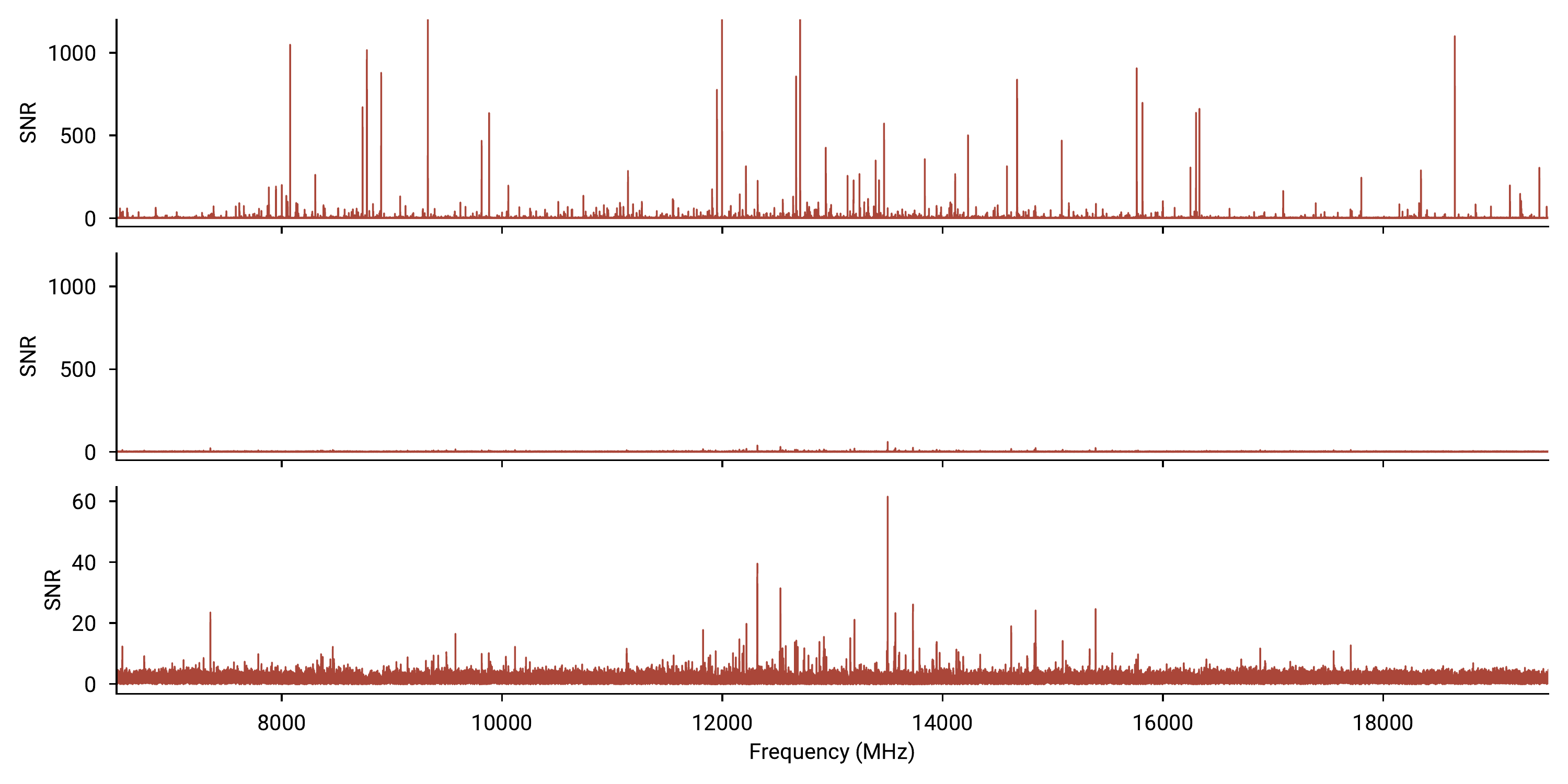}
    \caption{\scriptsize{Broadband spectrum of the discharge products  from a  dilute benzene/\ce{O2} mixture between 6.5 to 19.5\,GHz. The top panel shows the raw spectrum after approximately 18\,h of integration; the middle panel displays the identical spectrum but with all  assigned lines removed; the bottom panel is the same spectra shown in the middle panel but with the vertical axis expanded by a factor of 20. Unassigned features constitute approximately 14\% of the total number of observed lines (1158) and less than 4\% of the total intensity.  A total of 166 variants from 120 distinct product species were ultimately assigned in this spectrum.  Reprinted with permission from Ref.~\citenum{mccarthy:5170}. Copyright 2020 American Chemical Society.}}
    \label{fig:benzeneo2}
\end{figure*}

Perhaps the best illustration of these techniques and one of direct relevance to hydrocarbon chemistry is our recent analysis of a relatively simple system ---  an electrical discharge of benzene (\ce{C6H6}) alone \cite{lee:2408} or in combination with molecular oxygen or nitrogen \cite{mccarthy:5170}.  In total, 224 variants (including isotopic species and vibrationally excited states) from 160 unique product species  were identified between the three discharge mixtures.  Of these, approximately 50 product species have never been observed in the gas phase prior to this study.  Nearly 90\% of the nearly 3300 rotational features in the three spectra between 6.5 and 26\,GHz with a signal-to-noise ratio of six or higher were ultimately assigned (see Fig.~\ref{fig:benzeneo2}); these lines constitute approximately 97\% of the overall spectral intensity in this frequency range.  
 In these studies alone, we estimate the high-speed algorithm was used to derive preliminary rotational constants for upwards of 20 previously unknown molecules using no more than five double resonance linkages. Intriguingly, many of the product species are heavier in mass than the precursor benzene, indicating benzene growth occurs facilely under our experimental conditions. Furthermore, a significant number also deviate from the prolate limit, suggesting that we are better sampling all possible structures including those with three-dimensional structures, as evidenced from large negative inertial defects which is only possible when significant mass lies out-of-plane.  This work demonstrates two key findings: (i) the utility of microwave spectroscopy as a precision tool for complex mixture analysis, irrespective of whether the rotational spectrum of a product species is known \textit{a priori} or not; and (ii) the large discovery space that exists, provided the tools are in place to efficiently disentangle a mixture.
 
Once spectroscopic parameters of a molecule are established from experiment, the issue of paramount importance becomes molecule identification: determining the  elemental composition and unique structure of the carrier.  Arguably this step is the most challenging at present, particularly so as the rotational constants become progressively smaller, and the number of possible isomers grows rapidly. 
In our benzene analysis \cite{mccarthy:5170}, for example, compelling identifications for approximately half-a-dozen new molecules were not possible.  Nevertheless, there is no fundamental obstacle precluding an eventual identification; rather it would appear to be a matter of sustained effort in combination with full exploration of chemical space.
 
The recent astronomical detection of the two cyanocyclopentadienes \cite{mccarthy:inpress} shown in Fig.~\ref{fig:structures} came about as a direct result of this laboratory study.  Both isomers are prominent discharge products starting from benzene and molecular nitrogen, and although their rotational spectra were observed previously \cite{ford:326,sakaizumi:3903}, these measurements were done at considerably lower resolution (40-100\,kHz or 2-5\,ppm) compared to the present work (2\,kHz, or 0.1\,ppm).  As illustrated in Fig.~\ref{linewidth_comparison} the original line measurements are insufficient to make unambiguous identifications in the narrow-line sources like TMC-1.  The same laboratory study also provides a wealth of new or improved spectroscopic data for many hydrocarbon chains and rings and their nitrogen and oxygen derivatives at centimeter wavelengths.  It would be both disappointing and surprising if other molecules characterized in this work are not eventually found in TMC-1 or other molecular clouds.

It should be emphasized that the astronomical detections described here, like in the laboratory, occur in the gas phase.  In the molecular clouds such as TMC-1 where these species have been discovered, the ambient temperatures ($\sim$10\,K) mean that many collisions of larger molecules with dust grains result in condensation on the grain surface.  Studies of TMC-1 in particular have shown that relatively small variations in physical evolutionary state can have outsized effects on the gas-phase inventory \cite{Pratap:1997km}.  The presence of detectable quantities of large molecules, particularly aromatic species, thus implies gas-phase production routes as well as desorption mechanisms from grain surfaces are both efficient \cite{Marchione:2016hd}.  

\section{Inferring Formation Pathways by Isotopic Substitution\label{isotopic}}

Another strength of microwave investigations is the ability to use isotopic substitution to infer pathways that yield larger molecules from simpler precursors in our discharge sources.  Although discharges are frequently synonymous with non-selectivity and rapid isotopic scrambling, extensive $^{13}$C experiments on sulfur-terminated chains~\cite{mccarthy:ml07}, for example, establish that many are formed via fairly simple and direct reactions.  Evidence for $^{13}$C randomization was only found in a few instances, such as \ce{C3S}, which implies formation via cyclic intermediate.  With respect to cyanobenzene specifically, a direct and prompt pathway involving benzene and CN radical was demonstrated in our nozzle source in prior microwave study using \ce{CH3}$^{13}$CN in place of \ce{CH3CN} \cite{lee:2946}.

If benzene is predominately formed from the reaction between butadiene and the \ce{C2H} radical in our experiments, isotopic samples enriched in either D, $^{13}$C, or both should produce specific, isotopically-enriched products.  Use of \ce{CH2CDCHCH2} rather than butadiene in the reaction mixture with \ce{HC3N} (yielding both the CCH and CN radicals), for example, should yield singly-deuterated cyanobenzene.  Although a sample of pure \ce{CH2CDCHCH2} is not readily available, one enriched in this isotopic species (to the level of about 50\%) produced each deuterated species in comparable abundance. 
Use of high purity \ce{CH2CDCHCH2} should enable a more precise determination of abundance ratios, since purely statistical arguments predict a ratio of  2:2:1 for the \emph{ortho}:\emph{meta}:\emph{para} D positions, owing to the two unique substitution sites in singly-deuterated benzene that would yield either  \emph{ortho} or \emph{meta} $d_1$-cyanobenzene, but only a single site that would produce the \emph{para} species.   

A related isotopic experiment is to establish if the \ce{C2} unit of \ce{C2H} radical is conserved in ring formation.  If so, using \ce{H^{13}C2H} instead of \ce{HC2H} should produce doubly-substituted $^{13}$C cyanobenzene, but only isotopic species in which the two $^{13}$C atoms are adjacent to one anther in the ring skeleton. Isotopic studies that will lay the ground for this investigation --- detection of the unique doubly-substituted $^{13}$C species --- are currently underway in our laboratory. 

In terms of the aforementioned pyrrole studies, deuterated samples could also be exploited to investigate the formation pathway. Replacing \ce{NH3} with \ce{ND3} in the reaction with \ce{CH2(CH)2CH2} might be expect to yield $c$-\ce{C4H4N}-D selectively, as opposed to $c$-\ce{C4H3DN}-H.  If so, this result is consistent with prompt formation and one which implicates the ND (NH) radical.

In addition to mechanistic insight, it is worth noting that isotopic measurements conceivably could be used to distinguish between low and high temperature formation of PAHs in space.  For example, because isotopic fractionation sensitively depends on temperature, owing to differences to zero-point energies (e.g., HD), PAHs formed at low temperature may exhibit considerable D enhancement compared to high temperature routes which typify the outflow of evolved carbon stars.  Furthermore, the $^{13}$C/$^{12}$C ratios found in these evolved stars can substantially different from that in cold clouds, offering further evidence for a provenance. Much will depend on sensitivity, but the detection significance of benzonitrile in a deep ongoing survey of TMC-1 by spectral line stacking is in excess of 50$\sigma$ to date.  An analogous stack of its three unique $^{13}$C species has also yielded a tentative detection with low significance, however, additional observational data is needed to improve these values by factors of two or more before actionable insights could be contemplated.



    
    


\section{Future Work\label{futurework}}

Five aromatic or cyclic molecules, all CN-functionalized hydrocarbons, have been detected in a short period of time in a primordial gas cloud long known to harbor an unusual collection of unsaturated carbon chains.  These discoveries have come about because sensitive radio observations using the 100\,m GBT have revealed carbon chemistry of considerable richness and complexity just below the noise floor of previous spectral line surveys. The abundances of these rings are well in excess  of model predictions that well reproduce carbon chains -- in some cases by orders of magnitude --  implying that many other organic molecules may be present as well.  In light of these remarkable findings, there are potentially many  fruitful avenues of laboratory research to be pursued.  Some include:

\begin{enumerate}
    \item Systematic high-resolution rotational studies of many cyclic molecules and their derivatives, including polar PAHs, CN derivatives, and ones which incorporate heteroatoms such sulfur.  An important emphasis in terms of astronomical observations will be to determine the largest and most complex molecules in TMC-1, even if a deeper understanding of the underlying chemical complexity and its origin is lacking at present.   By closely coordinating laboratory measurements and astronomical observations it should be possible to develop a more complete picture of the chemical inventory in this and other molecular clouds.

    \item Laboratory detection of isomers of long, highly unsaturated carbon chains. Such studies are in direct response to the puzzling underabundance of \ce{HC11N} towards TMC-1\cite{loomis:inpress} despite strong lines of \ce{HC9N} and shorter cyanopolyynes there. The detection of matrix-isolated monocyclic \ce{C6} and \ce{C8} by infrared spectroscopy \cite{wang:6032} establishes that unsaturated monocyclic rings are indeed stable at modest sizes.  More recent optical spectroscopy of monocyclic carbon rings (\ce{C14}-\ce{C18}) by Maier and co-workers \cite{maier:369,boguslavskiy:127} lends further support to this argument.   Although indirect, ion chromatography studies of Bowers and others \cite{helden:1300,vonhelden:241,vonhelden:3835,gotts:217} suggest a wealth of structures, including large monocyclic rings, and bicyclic and tricyclic rings, starting at about 15 atoms. Direct observations of polar carbon clusters in this size range by microwave spectroscopy should provide  a better understanding of carbon cluster structure, how clusters form in the laboratory, and in what size range rings are more stable than chains.

    \item Testing mechanisms of PAH growth.  In the HACA mechanism,\cite{frenklach_formation_1989} PAHs are believed to form by repeated hydrogen abstraction followed by acetylene addition once benzene (or phenyl radical) is present, while other mechanisms invoke resonantly-stabilized radicals in growth.  Now that we can generate five and six-membered rings from acyclic precursors, a natural extension is to identify the pathways by which bicyclic rings are produced, for which indene formed from benzene and a source of carbon chains serves as a simple example.  Irrespective of the precise mechanism, if polar, reactive intermediates can be identified by high resolution microwave spectroscopy it should be possible to provide meaningful tests of PAH growth.  These types of studies will likely provide transition frequencies that will be of value for subsequent astronomical surveys.

\end{enumerate}

\section{Conclusions}

Time and time again nature has exhibited remarkable resilience in synthesizing molecules in what with appear to be inhospitable conditions to the chemical bond and organic chemistry.  Prior to the dawn of molecular astrophysics in the early to mid-1960's, the very existence of polyatomic molecules in space was viewed with great skepticism because, in retrospect, undue emphasis was placed on pathways, such as exposure to starlight, which might destroy or dissociate a molecule once formed.  As a result of this widely held misconception, the discovery of molecules in space came a surprise to many, but was soon rationalized by invoking a framework based on ion-molecule reactions. 

The skepticism regarding anions in space, albeit delayed by more than 40 years, followed a similar trajectory with absence of evidence slowing morphing into evidence of absence, driven again by unwarranted concerns about fragility to starlight.  We now know with considerable confidence that a wonderful assortment of familiar and quite exotic molecules are widespread in our galaxy and others.  There is every reason to expect that many more molecules can be found in space given advances in laboratory instrumentation and techniques, and the construction of increasingly powerful radio facilities. 

To this menagerie of interstellar molecules, we can finally add aromatics and small cyclic species that are far more familiar to the laboratory chemistry, and serve as the building blocks for many useful compounds on earth.  Five such rings have been discovered in rapid succession in TMC-1, and the detection of cyanobenzene elsewhere suggests aromaticity may be a common characteristic of many dark molecular clouds where star formation is thought to take place.  Perhaps not surprisingly given the history of the field, the abundance of these cyclic species is far greater than predicted from current chemical models which reproduce most carbon chains, implying the chemistry of two-dimensional carbon is favorable but poorly constrained at present. Although in its infancy, a closely coordinated and sustained theoretical, laboratory, and modeling effort is needed to understand the origin and fate of this unexpected aromatic content.   In doing so, these studies may shed light on acyclic-to-cyclic pathways and PAH growth at very low temperature and pressure, the role and importance of aromaticity in the interstellar medium, and the connection with much larger PAHs which are widely thought to be carriers of strong infrared emission features. 
With the past as a guide it would be unwise to underestimate nature,
 for it has proven to be far more elegant than we have insightful.

\bigskip
\noindent\textbf{Biographical information}

Dr. Michael C. McCarthy received a B.S. in Chemistry at the University of Alaska in 1986, and a PhD in Physical Chemistry from MIT in 1992. He then moved to the Harvard-Smithsonian Center for Astrophysics as a Center Fellow. In 1997, he joined the scientist staff being appointed the Yoram Avni Distinguished Research Astronomer. In 2014 he was named an Associate Director at the Center for Astrophysics where he leads the Atomic and Molecular Physics Division.  In 2020, he was appointed the Acting Deputy Director, a title he presently holds. His research interests include the spectroscopy of known and postulated astronomical carbon chains, carbon rings, and carbon clusters; the chemistry and physics of the interstellar medium; and molecular radio astronomy.

\includegraphics[scale=0.80]{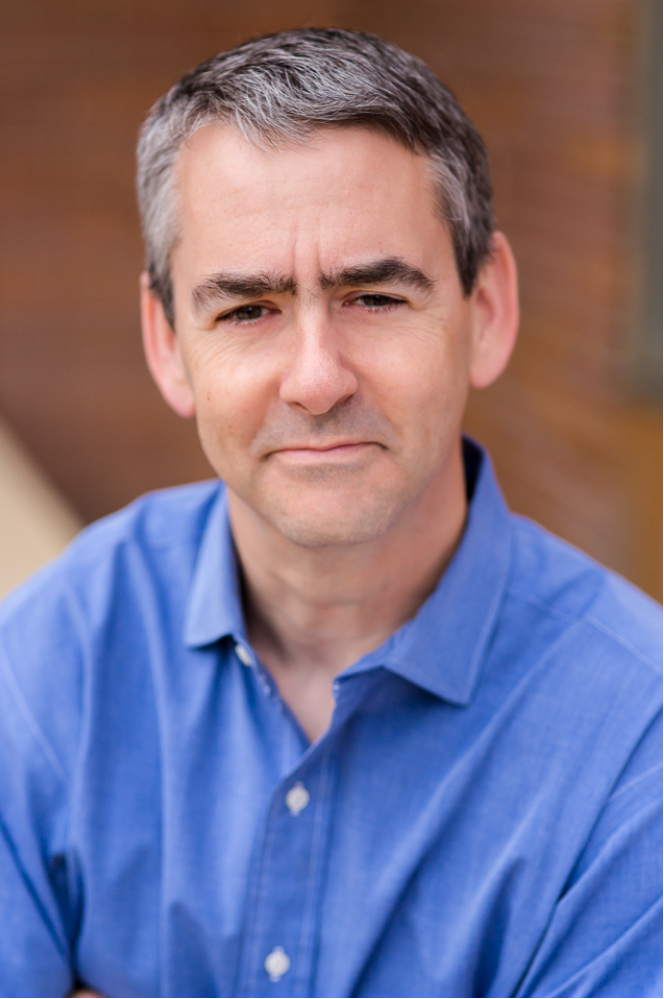}

Prof. Brett A. McGuire received his B.S. in Chemistry from the University of Illinois at Urbana-Champaign in 2009 and his Ph.D. in Physical Chemistry from the California Institute of Technology in 2014.  He was a National Radio Astronomy Observatory (NRAO) Jansky Fellow and then a NASA Hubble Fellow from 2014-2020 at the NRAO and the Center for Astrophysics | Harvard \& Smithsonian.  In 2020, he started as an Assistant Professor of Chemistry at the Massachusetts Institute of Technology.  Research in the McGuire Group uses the tools of physical chemistry, molecular spectroscopy, and observational astrophysics to understand how the chemical ingredients for life evolve with and help shape the formation of stars and planets.  He is particularly interested in the evolution of large, complex organic molecules.

\includegraphics[scale=0.30]{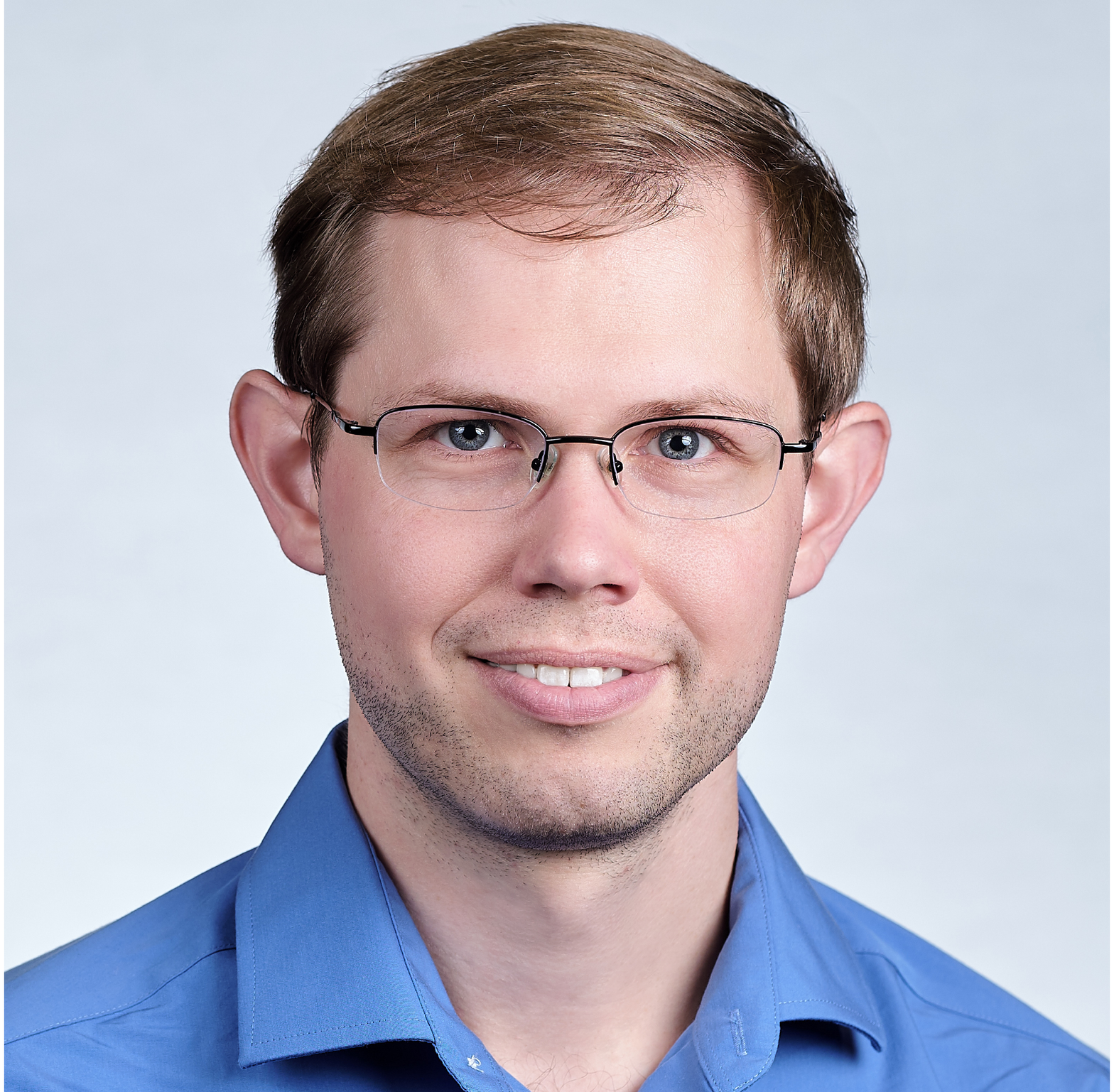}

\begin{acknowledgement}

The authors thank K.L.K. Lee  for help in preparing Fig.~\ref{fig_kinetics} , and J.P. Porterfield and members of the GOTHAM team for helpful comments.  Helpful comments of the anonymous reviewers are also acknowledged.   M.C.M.~acknowledges NSF grant AST-1908576 and NASA grant 80NSSC18K0396 for support. B.A.M. acknowledges prior support provided by NASA through Hubble Fellowship grant \#HST-HF2-51396 awarded by the Space Telescope Science Institute, which is operated by the Association of Universities for Research in Astronomy, Inc., for NASA, under contract NAS5-6555. 

\end{acknowledgement}


\providecommand{\latin}[1]{#1}
\makeatletter
\providecommand{\doi}
  {\begingroup\let\do\@makeother\dospecials
  \catcode`\{=1 \catcode`\}=2 \doi@aux}
\providecommand{\doi@aux}[1]{\endgroup\texttt{#1}}
\makeatother
\providecommand*\mcitethebibliography{\thebibliography}
\csname @ifundefined\endcsname{endmcitethebibliography}
  {\let\endmcitethebibliography\endthebibliography}{}

\end{document}